\documentstyle[epsfig]{article}

\newcommand{\be}{\begin{equation}}                                              
\newcommand{\ee}{\end{equation}}                                                
\newcommand{\half}{\frac{1}{2}}

\newcommand{\LCB}{\raisebox{-0.3ex}{\mbox{\LARGE$\left\{\right.$}}}
\newcommand{\RCB}{\raisebox{-0.3ex}{\mbox{\LARGE$\left.\right\}$}}}
\begin{document}
\title{
{\vspace{-3cm} \normalsize                                                      
\hfill \parbox{30mm}{DESY 97-153}}\\[15mm]
A Non-Chiral Extension of the Standard Model with                   \\
Mirror Fermions\footnote{\normalsize
Talk given at the workshop {\it Beyond the Standard Model V,}
Balholm, Norway, May 1997.}}

\author{ Istv\'an Montvay                                           \\
         Deutsches Elektronen-Synchrotron DESY                      \\
         Notkestr.\,85, D-22603 Hamburg, Germany}
\date{August, 1997}

\maketitle

\begin{abstract}
 The difficulties of defining chiral gauge theories non-perturbatively
 suggest a vector-like extension of the standard model with three
 mirror fermion families.
 Some phenomenological implications of such an extension are discussed.
\end{abstract}

%%%%%%%%%%%%%%%%%%%%%%%%%%%%%%%%%%%%%%%%%%%%%%%%%%%%%%%%%%%%%%%%%%%%%%%%
\section*{Introduction}
 The electroweak sector of the standard model is based on a chiral
 gauge theory with local chiral symmetry.
 Since the electroweak gauge couplings are weak, renormalized
 perturbation theory is an appropriate framework in most applications.
 Nevertheless, there are some questions where the non-abelian
 non-perturbative nature of the SU(2) coupling becomes relevant.
 For instance, this is the case for the electroweak phase transition
 where perturbation theory is plagued by severe infrared problems and
 also for some instanton induced multiparticle processes in high
 energy scattering.
 This makes a non-perturbative lattice formulation of chiral gauge
 theories not just a matter of principle but also an important
 practical question.

 It is a remarkable fact that, in contrast to vector-like gauge
 theories as QCD where the lattice formulation is easy and elegant,
 the non-perturbative lattice formulation of chiral gauge theories
 is very difficult, if not impossible.
 After many years of struggle one can conjecture that beyond the
 perturbative framework no chiral gauge theories exist!
 The assumptions, which are the basis for this conjecture, are:
\begin{itemize}
\item
 quantum field theories are defined as limits of reguralized cut-off
 theories;
\item
 the infinite cut-off limit exists and is independent of the choice of
 the regularization;
\item
 there exists an explicitly gauge invariant local action;
\item
 the formulation can be given by a Euclidean path integral satisfying
 reflection positivity, which implies unitarity after Wick rotation to
 Minkowski space.
\end{itemize}
 Under these, rather natural, assumptions all attempts for constructing
 a chiral gauge theory seem to fail.

 Relaxing some of these assumptions allows for lattice formulations
 which have a chance to give a consistent mathematical framework.
 Recently favorized examples are: the ``Rome-approach'' \cite{ROMA}
 where gauge fixing is assumed similarly to perturbation theory, the
 ``two-lattice approach'' \cite{TWO-LATTICE} where fermions are
 assumed to move on a much finer lattice than the gauge fields and
 the ``overlap formalism'' \cite{OVERLAP} where the path integral is
 abandoned and the  fermion determinant is defined by the overlap of
 ground states of five-dimensional auxiliary hamiltonians.
 Particularly striking is the impossibility to find an explicitly
 gauge invariant formulation: all these approaches have to break
 local gauge symmetry on the lattice and hope to be able to enforce
 gauge symmetry restoration in the continuum limit.
 This is in sharp contrast to vector-like gauge symmetries and to the
 very nature of local gauge invariance which arises due to the
 independence of internal symmetry orientations in different
 space-time points.

 A gauge invariant lattice formulation of an extension of the standard
 model becomes possible if one assumes the doubling of the physical
 fermion spectrum by mirror fermions \cite{GAUGEINV}.
 This implies that, besides the three ``left-handed'' fermion families,
 three ``right-handed'' mirror fermion families exist \cite{FAMILIES}.
 The existence of pairs of fermion-mirror-fermion families make the
 gauge symmetry ``vector-like'' and allows for explicit local gauge
 invariance.

 The mirror fermion families have not been experimentally observed up
 to now and may only exist if their masses are high enough and their
 mixings with the observed light fermions are small enough.
 Some possible deviations from the standard model, observed in recent
 years, might have to do with direct or indirect effects of the mirror
 fermions.
 Examples are possible small violations of the universality of gauge
 couplings \cite{UNIVERSAL} and/or the excess of high $Q^2$ events at
 HERA, which can be caused by a new strong interaction of mirror
 fermions at the TeV-scale \cite{STRONG}.
 In spite of some possible positive signals, an overall detailed study
 of the phenomenological viability of the vector-like extension of the
 standard model is still missing.
 Nevertheless, first investigations have already been carried out at
 the one-loop perturbative level \cite{ONE-LOOP}.
 In the present talk I shall point out a few recently discussed
 possibilities within the mirror fermion framework.

%%%%%%%%%%%%%%%%%%%%%%%%%%%%%%%%%%%%%%%%%%%%%%%%%%%%%%%%%%%%%%%%%%%%%%%%
\section*{Mirror fermions: mixing schemes}
 In order to illustrate the possible mixing between fermions and their
 mirror fermion partners, let us consider the mass matrix of a
 fermion-mirror-fermion pair on the chiral basis
 $(\overline{\psi}_R,\overline{\psi}_L,
 \overline{\chi}_R,\overline{\chi}_L)
 \otimes (\psi_L,\psi_R,\chi_L,\chi_R)$:
\be \label{eq01}
M = \left( \begin{array}{cccc}                                                  
\mu_\psi  &  0  &  \mu_R  &  0       \\                                         
0  &  \mu_\psi  &  0  &  \mu_L       \\                                         
\mu_L  &  0  &  \mu_\chi  &  0       \\                                         
0  &  \mu_R  &  0  &  \mu_\chi                                                  
\end{array} \right) \ .                                                         
\ee                                                                             
 Here $\mu_{(L,R)}$ are the fermion-mirror-fermion mixing mass
 parameters, and the diagonal elements are produced by spontaneous
 symmetry breaking:
\be \label{eq02}
 \mu_\psi=G_\psi v \ , \hspace{2em} \mu_\chi=G_\chi v \ ,
\ee                                                                             
 with the Yukawa-couplings $G_{\psi}$, $G_{\chi}$ and the vacuum
 expectation value of the Higgs scalar field $v$.
                                                                                
 For $\mu_R \ne \mu_L$ the mass matrix $M$ is not symmetric, hence one
 has to diagonalize $M^T M$ by $O^T_{(LR)} M^T M O_{(LR)}$, and
 $M M^T$ by $O^T_{(RL)} M M^T O_{(RL)}$, where
\be \label{eq03}
O_{(LR)} = \left( \begin{array}{cccc}                                           
\cos\alpha_L  &  0  &  \sin\alpha_L  &  0   \\                                  
0  &  \cos\alpha_R  &  0  &  \sin\alpha_R   \\                                  
-\sin\alpha_L  &  0  &  \cos\alpha_L  &  0  \\                                  
0  &  -\sin\alpha_R  &  0  &  \cos\alpha_R                                      
\end{array} \right) \  ,                                                        
\ee                                                                             
 and $O_{(RL)}$ is obtained by exchanging the indices
 $R \leftrightarrow L$.
 The rotation angles of the left-handed, respectively, right-handed             
 components satisfy
\be \label{eq04}
\tan(2\alpha_L) = \frac{2(\mu_\chi \mu_L + \mu_\psi \mu_R)}                     
{\mu_\chi^2 + \mu_R^2 - \mu_\psi^2 - \mu_L^2} \ ,
\hspace{2em}
\tan(2\alpha_R) = \frac{2(\mu_\chi \mu_R + \mu_\psi \mu_L)}                     
{\mu_\chi^2 + \mu_L^2 - \mu_\psi^2 - \mu_R^2} \ .                               
\ee                                                                             
 The two (positive) mass-squared eigenvalues are given by
$$
\mu_{1,2}^2 = \half \LCB                                                     
\mu_\chi^2 + \mu_\psi^2 + \mu_L^2 + \mu_R^2 
\mp \left[ (\mu_\chi^2 - \mu_\psi^2)^2 + (\mu_L^2 - \mu_R^2)^2
\right.
$$
\be \label{eq05}
\left.
+(\mu_L^2 - \mu_R^2)^2+ 2(\mu_\chi^2 + \mu_\psi^2) (\mu_L^2 + \mu_R^2)
+8\mu_\chi \mu_\psi \mu_L \mu_R \right]^\half \RCB \ .
\ee
 For $\mu_\psi,\mu_L,\mu_R \ll \mu_\chi$ there is a light state with
 $\mu_{1}={\cal O}(\mu_\psi,\mu_L,\mu_R)$ and a heavy state with
 $\mu_{2}={\cal O}(\mu_\chi)$.
 In general, both the light and heavy states are mixtures of the
 original fermion and mirror fermion.

 In case of three mirror pairs of fermion families the diagonalization
 of the mass matrix is in principle similar but, of course, more                
 complicated \cite{FAMILIES}.
 The non-observation of mirror fermion pair production at LEP implies
 that the masses of the mirror fermions, including massive mirror
 neutrinos, have to be above $\simeq$90 GeV.
 The mixing angles among fermions and mirror fermions have to be small,
 in order to avoid an excessive violation of universality in $W$- and
 $Z$-boson couplings.
 The limits on the mixing angles are stronger for the first fermion
 family than for other families, and stronger for leptons than for
 quarks \cite{UNIVERSAL}.

%%%%%%%%%%%%%%%%%%%%%%%%%%%%%%%%%%%%%%%%%%%%%%%%%%%%%%%%%%%%%%%%%%%%%%%%
\section*{Leptoquarks and new strong interactions}
 If the mirror families are very heavy then the renormalization of the
 Yukawa couplings implies strong interactions at a relatively nearby
 energy scale above the electroweak scale.
 (See the bounds on mirror fermion masses following from the requirement
 of perturbative unification \cite{BOUNDS}).

\begin{figure}[ht!]
\centerline{\epsfig{file=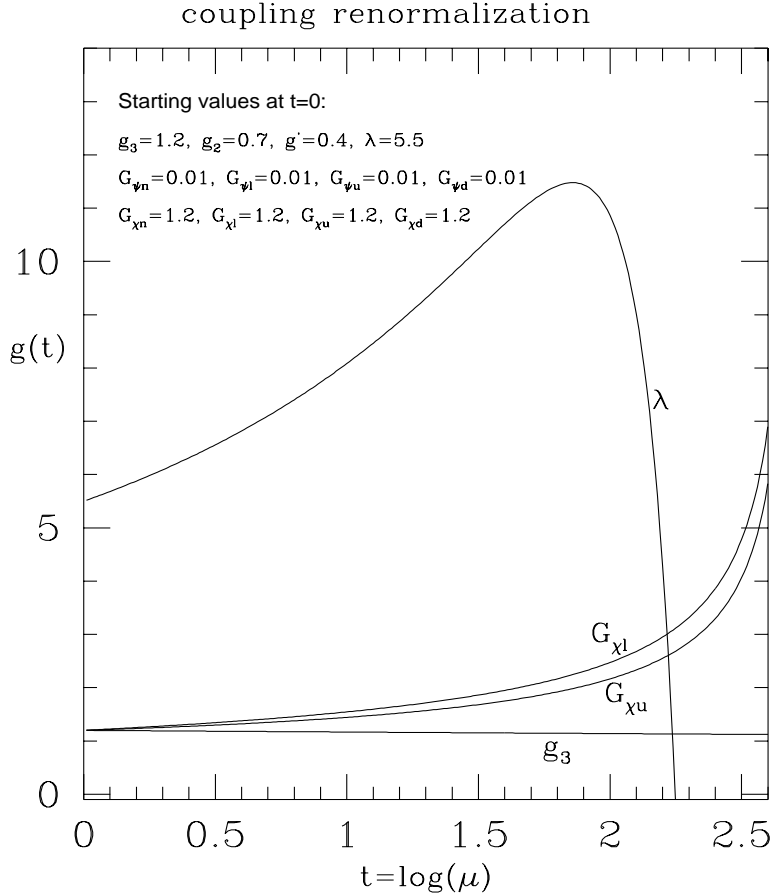,
           width=11.0cm,height=12.0cm,
           bbllx=80pt,bblly=150pt,bburx=590pt,bbury=730pt,
           angle=0}}
\vspace{10pt}
\caption{Numerical solution of the renormalization group equations for
 gauge-, quartic- and Yukawa-couplings.
 The initial values are shown in the left upper corner.}
\label{fig1}
\end{figure}

 Let us denote, as usual, the gauge couplings for SU(3), SU(2) and U(1),
 respectively, by $g_3$, $g_2$ and $g^\prime$.
 The Yukawa couplings for light fermions are $G_\psi ..$ and for mirror
 fermions $G_\chi ..$, whereas the quartic scalar coupling is denoted 
 by $\lambda$.
 The renormalization group equations for these couplings are, for
 simplicity, at one-loop level with equal Yukawa couplings for 
 $\nu$-, $l$-, $u$- and $d$-type fermions:
$$
\frac{dg_3}{dt} = -\frac{3g_3^3}{16\pi^2} \ ,
\hspace{2em}
\frac{dg_2}{dt} = \frac{5g_2^3}{96\pi^2} \ ,
\hspace{2em}
\frac{dg^\prime}{dt} = \frac{81g^{\prime 3}}{96\pi^2} \ ,
$$
$$
\frac{dG_{\psi\nu}}{dt} = \frac{G_{\psi\nu}}{16\pi^2}
\left\{ -{9 \over 4}g_2^2 -{3 \over 4}g^{\prime 2} 
+{9 \over 2}G_{\psi\nu}^2 +{3 \over 2}G_{\psi l}^2
\right.
$$
$$
\left.
+3\left( G_{\chi\nu}^2+G_{\chi l}^2 \right) 
+9\left( G_{\psi u}^2+G_{\psi d}^2+G_{\chi u}^2+G_{\chi d}^2 \right) 
\right\} \ ,
$$
$$
\frac{dG_{\psi l}}{dt} = \frac{G_{\psi l}}{16\pi^2}
\left\{ -{9 \over 4}g_2^2 -{15 \over 4}g^{\prime 2} 
+{9 \over 2}G_{\psi l}^2 +{3 \over 2}G_{\psi\nu}^2
\right.
$$
$$
\left.
+3\left( G_{\chi\nu}^2+G_{\chi l}^2 \right) 
+9\left( G_{\psi u}^2+G_{\psi d}^2+G_{\chi u}^2+G_{\chi d}^2 \right) 
\right\} \ ,
$$
$$
\frac{dG_{\psi u}}{dt} = \frac{G_{\psi u}}{16\pi^2}
\left\{ -8g_3^2 -{9 \over 4}g_2^2 -{17 \over 2}g^{\prime 2} 
+{21 \over 2}G_{\psi u}^2 +{15 \over 2}G_{\psi d}^2
\right.
$$
$$
\left.
+3\left( G_{\psi\nu}^2+G_{\psi l}^2+G_{\chi\nu}^2+G_{\chi l}^2 \right) 
+9\left( G_{\chi u}^2+G_{\chi d}^2 \right) \right\} \ ,
$$
$$
\frac{dG_{\psi d}}{dt} = \frac{G_{\psi d}}{16\pi^2}
\left\{ -8g_3^2 -{9 \over 4}g_2^2 -{5 \over 12}g^{\prime 2} 
+{21 \over 2}G_{\psi d}^2 +{15 \over 2}G_{\psi u}^2
\right.
$$
$$
\left.
+3\left( G_{\psi\nu}^2+G_{\psi l}^2+G_{\chi\nu}^2+G_{\chi l}^2 \right) 
+9\left( G_{\chi u}^2+G_{\chi d}^2 \right) \right\} \ ,
$$
$$
\frac{dG_{\chi \ldots}}{dt} = 
\left\{ \psi \leftrightarrow \chi \right\} \ ,
$$
$$
\frac{d\lambda}{dt} = \frac{1}{16\pi^2}
\left\{ 4\lambda^2 -\lambda\left( g_2^2+3g^{\prime 2} \right) 
+{27 \over 4}g_2^4 
+{9 \over 2}g_2^2g^{\prime 2} +{9 \over 4}g^{\prime 4}
\right.
$$
$$
+12\lambda\left( G_{\psi\nu}^2+G_{\psi l}^2
+G_{\chi\nu}^2+G_{\chi l}^2 \right) 
+36\lambda\left( G_{\psi u}^2+G_{\psi d}^2
+G_{\chi u}^2+G_{\chi d}^2 \right)
$$
\be \label{eq06}
\left.
-36\left( G_{\psi\nu}^4+G_{\psi l}^4
+G_{\chi\nu}^4+G_{\chi l}^4 \right)
-108\left( G_{\psi u}^4+G_{\psi d}^4
+G_{\chi u}^4+G_{\chi d}^4 \right) \right\} \ .
\ee                                                                             
 The variable $t$ is defined as usual by $t \equiv \log\mu$.
 The assumed degeneracy of $\nu$-, $l$-, $u$- and $d$-type fermions
 in the three mirror pairs of fermion families does not change the
 qualitative behaviour, because the Yukawa-couplings are in any case
 dominated by the heavy mirror fermions.
 Not even the separate inclusion of a heavy top quark has a strong
 influence.
 A typical behaviour of the couplings as a function of $t$ is shown by
 figure \ref{fig1}.
%%%%%%%%%%%%%%%%%%%%%%%%%%%%%%%%%%%%%%%%%%%%%%%%%%%%%%%%%%%%%%%%%%%%%%%%
%%%%%%%%%%%%%%%%%%%%%%%%%%%%%%%%%%%%%%%%%%%%%%%%%%%%%%%%%%%%%%%%%%%%%%%%
 As one can see, for the given initial conditions corresponding to
 mirror fermion masses of about 200 GeV, the quartic and Yukawa
 couplings diverge at an energy scale about a factor of 10 higher than
 the initial (electroweak) scale.

 The divergence of the Higgs scalar couplings signals the existence of
 new strong interactions in the TeV energy range.
 A simple scheme within the vector-like extension of the standard model,
 based on a strong U(1) interaction, has been considered recently in
 \cite{STRONG}.
 A strong U(1) interaction has been first proposed for the explanation
 of the high $Q^2$ HERA anomaly in ref.~\cite{STRONG-U1}.
 Scenarios with strongly interacting non-trivial U(1) gauge theories
 are supported by recent lattice investigations showing non-trivial
 scaling behaviour in the continuum limit \cite{LATTICE-U1}.
 (For further references see this paper.)

 Within the vector-like extension of the standard model the nature of
 the new (vector-like) strong interactions, which eventually produce
 leptoquark bound states, is quite arbitrary.
 Besides U(1), another interesting possibility is to assume that the
 strong interaction at the TeV scale involves the $\rm SU(2)_R$ gauge 
 group, which is originally not gauged.
 (Note that $\rm SU(2)_R$ with three mirror pairs of families is not
 asymptotically free.
 Asymptotic freedom holds if the $\rm SU(2)_R$'s for the three
 families are gauged separately as $\rm SU(2)_R^{\otimes 3}$.)
 In this case leptoquarks can appear both in neutral current and
 charged current $e^+p$ processes.
 In fact, assuming strong OZI-rule for the couplings of leptoquarks
 to their constituents (as in \cite{STRONG}), the dominant effective
 four-fermion couplings relevant for $e^+d$-scattering are:
$$
{\cal L} \simeq \alpha_L^{(l)2} (\overline{e}_L d_R)
\left[ (\overline{u}_R \nu_L)+(\overline{d}_R e_L) \right]
$$
\be \label{eq07}
+\alpha_L^{(q)2} (\overline{e}_R d_L)
\left[ (\overline{u}_L \nu_R)+(\overline{d}_L e_R) \right] 
+ {\cal O}(\alpha_L^{(l)}\alpha_L^{(q)}) \ .
\ee                                                                             
 The omitted off-diagonal terms proportional to
 $\alpha_L^{(l)}\alpha_L^{(q)}$ are strongly constrained by low energy
 data \cite{CC}.
 Assuming that the product $\alpha_L^{(l)}\alpha_L^{(q)}$ is
 very small and one of the diagonal terms shown in eq.~(\ref{eq07})
 dominates, the low energy constraints are avoided.
 The consequence of eq.~(\ref{eq07}) is that the decay branching ratio
 of the letoquark in $e^+d$-scattering to neutral current and charged
 current final states is roughly the same.

%%%%%%%%%%%%%%%%%%%%%%%%%%%%%%%%%%%%%%%%%%%%%%%%%%%%%%%%%%%%%%%%%%%%%%%%

\end{document}